\documentclass[aps,twocolumn,floatfix,superscriptaddress,pre,reprint]{revtex4}
\usepackage{amssymb,amsmath,graphicx,times}
\usepackage{mathptmx}
\usepackage[usenames]{color}

\usepackage{amsfonts,amsthm,bbm} % Math packages
\usepackage{amsmath}
\usepackage{amstext}
\usepackage{amsfonts}
\usepackage{amssymb,amscd}
\usepackage{amsopn}
\usepackage{graphicx}
\usepackage{physics}
\usepackage{mathtools}
\usepackage{lscape}
\usepackage{multirow}
\usepackage{diagbox}
\usepackage{color}
\usepackage[dvipsnames]{xcolor}
\usepackage{hyperref}
\usepackage{tensor}
\usepackage{bm,bbm}
\usepackage{epsfig}
\usepackage[caption=false]{subfig}
\usepackage{todonotes}

\newcommand{\diag}{\mathrm{diag}}
\newcommand{\1}{\mathbb{I}}
\newcommand{\Id}{\1}

\newtheorem{proposition}{Proposition}

\begin{document}

\title{Matrix logistic map: fractal spectral distributions and transfer of chaos}

\author{{\L}ukasz Pawela}
\affiliation{Institute of Theoretical and Applied Informatics, Polish Academy
of Sciences, ul. Ba{\l}tycka 5, 44-100 Gliwice, Poland}
\author{Karol {\.Z}yczkowski}
\affiliation{Institute of Physics, Jagiellonian University, ul. {\L}ojasiewicza
11, 30--348 Krak\'ow, Poland} 
\affiliation{Center for Theoretical Physics,
Polish Academy of Sciences, Al. Lotnik\'{o}w 32/46, 02-668 Warszawa, Poland}

\date{August 13, 2025}

\begin{abstract}
\noindent
The standard logistic map, $x'=ax(1-x)$, serves as a paradigmatic model to
demonstrate how apparently simple nonlinear equations lead to complex and
chaotic dynamics. In this work we introduce and investigate its matrix analogue
defined for an arbitrary matrix $X$ of a given order $N$. We show that for an
arbitrary initial ensemble of  Hermitian random matrices with a continuous level
density supported on the interval $[0,1]$, the asymptotic level density
converges to the invariant measure of the logistic map. Depending on the
parameter $a$ the constructed measure may be either singular, fractal or
described by a continuous density. In a wider class of the map, 
the multiplication by a scalar logistic parameter $a$ is replaced by 
transforming $aX({\mathbbm I}-X)$ 
into $BX({\mathbbm I}-X)B^{\dagger}$, 
where  $A=BB^{\dagger}$  is a fixed
positive matrix of order $N$. This approach generalizes the known model of
coupled logistic maps, and allows us to study the transition to chaos in complex
networks and multidimensional systems. In particular, by associating the matrix $B$
with a given graph we demonstrate the gradual transfer of chaos between
subsystems corresponding to vertices of a graph and coupled according to its
edges.
\end{abstract}

\maketitle

{\bf Complicated nonlinear dynamics can often be explored through the analysis
of simple maps defined on an interval. Motivated by recent advances in quantum
algorithms, we extend this idea to study analogous maps acting on spaces of
matrices. We introduce a matrix generalization of the standard logistic map
which be viewed as a generalization of the model of coupled maps with topology
of coupling between nodes determined by a specified directed graph. Our approach
leads to novel ensembles of random matrices with fractal spectral measures.}

\section{Introduction}
The theory of dynamical systems is used to explain complex time evolution
characteristic for numerous problems in physics and beyond \cite{St00}. A
particularly simple 1D system, called a {\sl logistic map},
\begin{equation}
x_{n+1}=f_a(x_n) \ := \ a x_n (1-x_n)
\label{log1}
\end{equation}
for $x\in [0,1]$, proved to be useful to understand various scenarios of the
transition from regular to chaotic dynamics \cite{MA76, AD06}. Investigation of
changes of the dynamics with the logistic parameter $a\in (0,4]$ led to several
seminal discoveries, such as the phenomenon of period doubling, the Feigenbaum
universality typical to parabolic extrema of the map \cite{Fe78}, and the
observation of the Sharkovskii order of periodic orbits \cite{Sh64, LM94}, which
culminates in the known statement: {\sl Period three implies chaos} \cite{LY75}.
Logistic map and other chaotic systems were used to design encryption schemes
\cite{KJSP98,KJ01}.

Various models of coupled logistic maps, popularized by Kaneko  \cite{Ka93},
offer a much wider range of dynamical behavior than the single map
\cite{MM03,MG18} and allow one to analyze interactions between several chaotic
subsystems and to study possible synchronization between them \cite{ABV06}.
Dynamical systems of coupled map lattices found numerous applications in
statistical physics to study spacetime chaos \cite{BS88}, but also in
theoretical chemistry to model chemical reactions \cite{FR88} and in
evolutionary biology to study population dynamics \cite{LL95,KF98}.

Another class of extensions of the standard logistic map was proposed by
Navickas et al. \cite{NSVR11,NRVS12}. In a recent paper \cite{SNR21} they
considered an arbitrary dimension $N$ and studying logistic equation for
non-Hermitian matrices observed effects of explosive divergence of the iterative
process. The early Lithuanian papers inspired another attempt to investigate
dynamics of logistic map defined on matrices of order two \cite{Pre12}.

In this contribution we investigate a powerful class of nonlinear dynamical
systems which include the logistic map that acts on matrices of an arbitrary
order $N$. The map defined on Hermitian positive matrices, $H^{\dagger}=H\ge 0$,
with bounded operator norm,  $||H|| \le 1$, can be related to models of
nonlinear quantum evolution of density matrices \cite{Te99}, in which the
quadratic term, $\rho^2$, corresponds to quantum measurements performed on two
copies \cite{BPTC09} of the same quantum state $\rho$. 

Other quantum applications include nonlinear transformations, in which
individual entries of the density matrix are squared \cite{BPHG98},
measurement-based nonlinear rotations of the Bloch sphere \cite{HS01}, more
general measurement-induced transformations  \cite{KK18,PKJK22} and
investigations of reduced quantum dynamics \cite{KKK+22}. Recent interest in
dynamics in the space of matrices is motivated by the advent of quantum
algorithms applied to solve algebraic problems \cite{HHL09,MRTC21,MF24}.

The aim of  this work is to initiate further studies on a broad spectrum of
dynamics corresponding to the matrix logistic map acting on arbitrary matrices
of a finite dimension $N$ with a bounded norm. In this way we establish a link
between logistic map and the  theory  of random matrices and study dynamics in
the space of ensembles of matrices. In particular, a class of ensembles of
Hermitian random matrices of any fixed dimension $N$ with eigenvalues
distributed according to fractal measures is constructed. Other assumptions lead
to ensembles of non-Hermitian matrices with eigenvalues belonging to a single
ring in the complex plane \cite{FZ97,GKZ11} and singular values distributed
according to fractal probability measures. Making use of the theory of free
random matrices \cite{Vo89} we obtain in some cases exact expressions for the
spectral density valid in the asymptotic case $N\to \infty$.

In the general form of the model studied the logistic parameter $a$ is also
replaced by a fixed positive matrix $A$. On the one hand, investigations of
nonlinear maps acting in the space of matrices are relevant in study of the
transition to chaos for coupled nonlinear classical systems. On the other hand,
they are useful for various models of nonlinear quantum evolution of density
matrices, corresponding to interacting quantum systems.

Key applications of the model concern the matrix logistic  
map analogous to (\ref{log1}), with the scalar parameter $a$ replaced by a
positive matrix $A=BB^{\dagger} \ge 0$ such that $A\le 4 {\mathbb I}_N$. Such a
matrix model proves to be useful to investigate the transfer of chaos in a
composed dynamical system with interactions determined by the graph associated
to the matrix $B$.

\section{Standard logistic map}
Dynamics of the map (\ref{log1}) is determined by the logistic parameter $a \in
[0,4]$. For $a<3$ there exists a single fixed point of the map, while at $a=3$ a
two-cycle appears. A sequence of period doubling effects leads to the onset of
chaos at $a=a_c \approx 3.569$, for which the Lyapunov exponent $\Lambda$
becomes positive as almost all initial conditions do not lead to periodic
dynamics. The  value $a_* \approx 3.828$ corresponds to  a periodic window, in
which one observes mixed dynamics composed of chaotic trajectories and regular
behavior. Apart of periodic windows the invariant measure  $\mu_a$ typically
exhibits a fractal structure.
 
\begin{figure}[!h]
\centering
\includegraphics[width=\columnwidth]{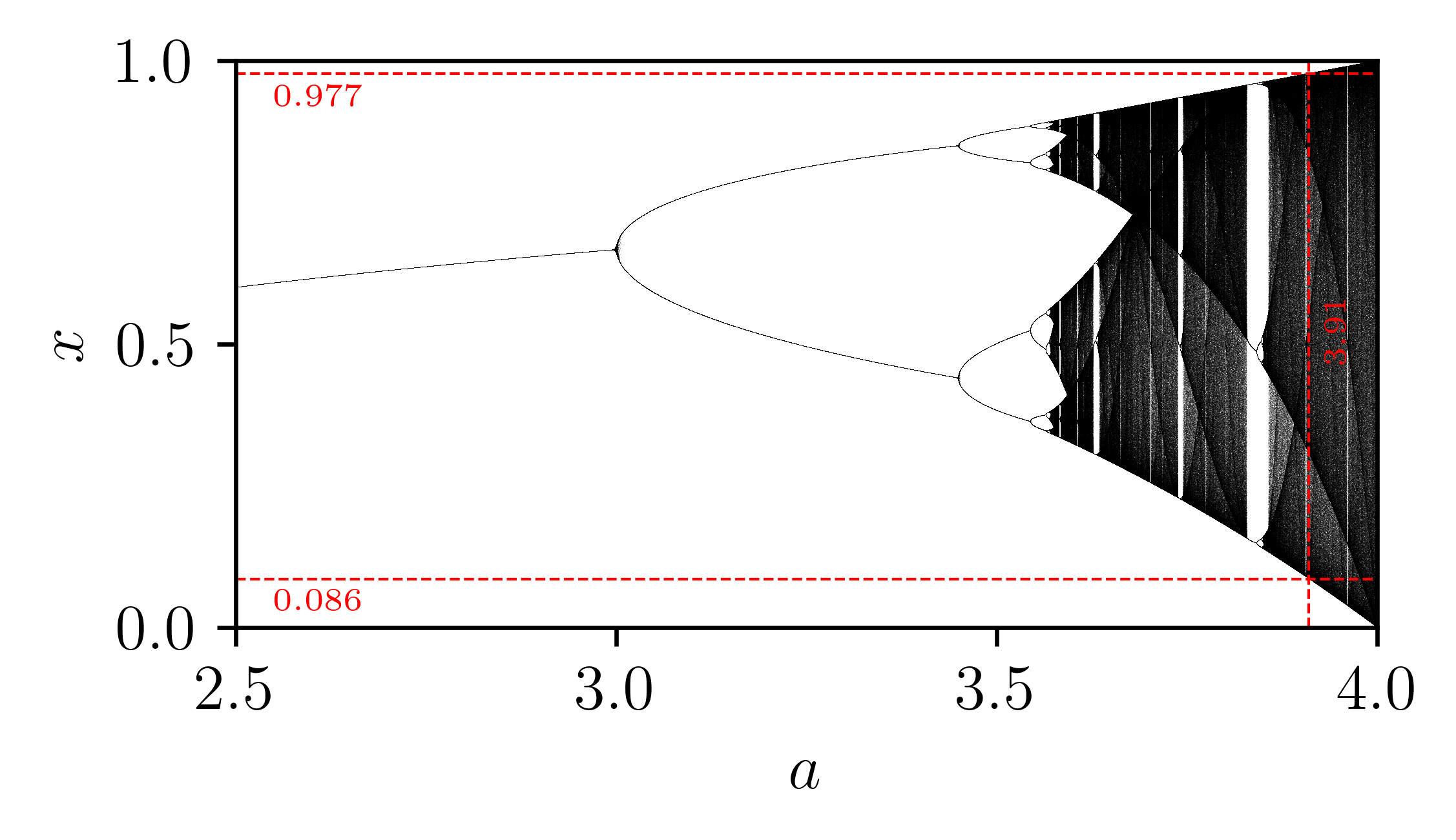}\\
\includegraphics[width=\columnwidth]{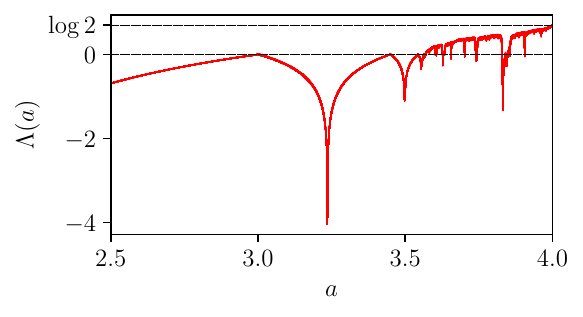}
\caption{(top) Bifurcation diagram for  logistic map for control parameter
$a\in [2.5,4.0]$ and (bottom) Lyapunov exponent $\Lambda(a)$. Observe
bifurcations, period doubling,  $2^k$--cycles, periodic windows, Sharkovskii
order, three-implies-chaos rule and transition to full chaos at $a=4$. Values of
the parameter $a$ marked by short vertical lines were used to obtain results on
the matrix version of the map.}
\label{fig:bifur}
\end{figure}

For convenience of the reader we reproduce in Fig.\ref{fig:bifur} the known
bifurcation diagram for the map (\ref{log1}), with the corresponding Lyapunov
exponent $\Lambda(a)$. For a given value of $a$ one can identify  at the
vertical axis the interval $[x_{\rm min}, x_{max}]$, in which the invariant
measure $\mu_a$ is supported. Marked values of the logistic parameter $a$
represent exemplary cases, for which the generalized matrix maps are analyzed.

 In the limiting case, $a=4$, the motion becomes fully chaotic on the interval
$I=[0,1]$ with the Lyapunov exponent  $\Lambda=\log 2$. A nonlinear change of
the variables allows one to bring this equation to the tent map, which implies
that the invariant measure of the corresponding transition operator is given by
the so called {\sl arc-sin law}, 
\begin{equation}
\label{arcsin}
 \mu_4^*(x)=\frac{1}{\pi \sqrt{x(1-x)} } 
\end{equation}
 -- see e.g. \cite{Ott}.
 
\section{Matrix logistic map: Hermitian matrices}
To extend the logistic map (\ref{log1}) for the space of Hermitian matrices of
order $N$, we need to bound the support of spectrum of a given positive
Hermitian matrix, $H=H^{\dagger}\ge 0$. Rescaling the matrix by its operator
norm, $||H||_{\rm op}$, equal to the largest eigenvalue $\lambda_{\rm max}$, we
obtain the matrix $H_0=H/\lambda_{\rm max}$ with spectrum contained in the unit
interval, $I=[0,1]$. This matrix can be used as a starting point for  the
following {\sl matrix logistic map},
\begin{equation}
H_{t+1}=f_a(H_t) \ := \ a H_{t} ({\mathbb I} -H_t),
\label{log2}
\end{equation}
which acts in the set Hermitian matrices with spectrum supported on the unit
interval.

Any Hermitian matrix can be diagonalized by a unitary transformation,
$H_t=UD_tU^{\dagger}$, where $D_t$ denotes a diagonal matrix with eigenvalues of
$H_t$ at the diagonal. Inserting this form into  Eq. (\ref{log2}) we see that
the matrix $U$ of eigenvectors of $H_0$ is preserved, so the logistic dynamics
concerns the eigenvalues only,
\begin{equation}
D_{t+1}=f(D_t) \ := \ a D_{t} ({\mathbb I} -D_t) = a (D_t-D_t^2) .
\label{log3}
\end{equation}

Thus, the matrix logistic map (\ref{log2}) applied to a given matrix $H_0$ of
order $N$ can be interpreted as a parallel iteration of $N$ individual initial
points, corresponding to the eigenvalues of $H_0$ by the scalar map
(\ref{log1}). As any continuous measure $\nu$ on the interval $I$ iterated by
the Markov operator $M_a$, associated with the logistic map (\ref{log1}), is
known to converge to the limiting invariant measure $\mu_a^*=M_a\mu_a^*$, see
\cite{LM94}. While there may be several invariant measures, only one is
physical. The physical measure is an ergodic measure that characterizes the
typical behavior of almost every initial condition, as observed with respect to
the Lebesgue measure. This concept of a physical measure, which aligns with the
SRB (Sinai-Ruelle-Bowen)~\cite{Yo02} measure, provides a robust description of
the system's dynamics and has been extensively studied. Thus, we arrive at the
following statement.

\begin{proposition}\label{prop:1}
Consider an ensemble of random Hermitian matrices of order $N$ with spectral
density given by any continuous function supported in the unit interval.
Iterating its elements $t$ times by the logistic matrix map (\ref{log2}) 
one obtains asymptotically an ensemble with spectral density converging in the
limit $t \to \infty$ to the invariant measure $\mu_a^*$ of the logistic map
(\ref{log1}) with parameter $a$.
\end{proposition}

\begin{figure}[h]
\centering
\subfloat{\includegraphics[width=\columnwidth]{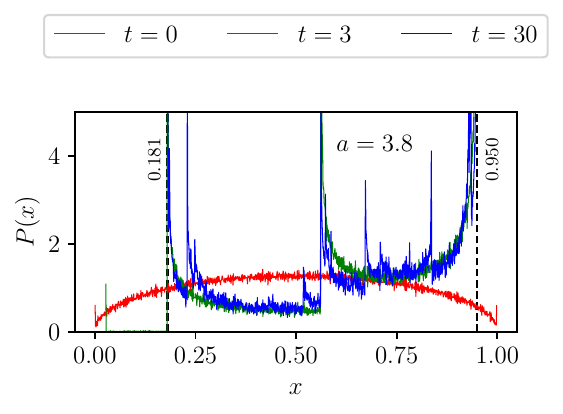}\label{fig:iterate-gue}}\\
\subfloat{\includegraphics[width=\columnwidth]{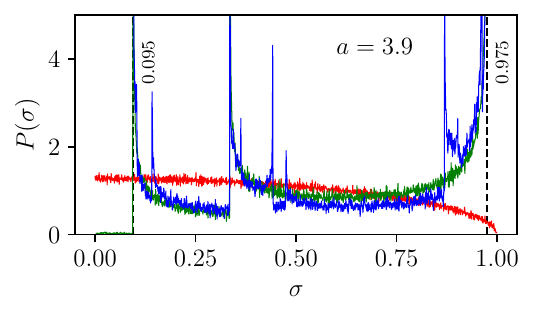}\label{fig:iterate-ginibre}}
\caption{Spectral distribution $P(y)$ of the rescaled GUE ensemble of size
$N=1000$ iterated by the logistic map for $a=3.8$ at times $t=0$  -- initial
density equivalent to a shifted Wigner semicircle (red), $t=3$ iterations
(green), and $t=30$ iterations approximating the invariant measure $\mu_{3.8}^*$
(blue). b) Similar plot for dynamics of distribution $P(x)$ of singular values
of the rescaled random Ginibre matrices, iterated by the logistic map with
parameter $a=3.9$. Initial  distribution of  Marchenko--Pastur converges to
fractal measure $\mu_{3.9}^*$. The black dashed lines mark the support of the
fractal measures $\mu_a^*$. In this context, \emph{fractal measure} refers to
the physical measure that describes the system's long-term behavior.}

\label{fig:herm38}
\end{figure} 

Selecting values of the logistic parameter $a$ such that the invariant measure
$\mu_a^*$ is known to display fractal properties we construct an ensemble of
random Hermitian matrices with such a spectral distribution. Fast convergence of
the initial semicircle density, corresponding to the Gaussian unitary ensemble
(GUE) \cite{Me04}, to a fractal measure $\mu_a^*$  determined by the logistic
parameter $a$ is visualized in Fig.~\ref{fig:iterate-gue}. In the limiting case
of $a=4$, corresponding to the chaotic dynamics in the entire interval, the
invariant measure becomes continuous and is given by the arcsine law (\ref{arcsin}).

Note also that iterating a single random matrix $H$ of size $N$,  the spectrum
of the matrix $f_a^t(H)$ is described, in the limit of large matrix dimension,
$N\to \infty$, by the invariant measure $\mu_a^*$.

\section{Matrix logistic map:   non-Hermitian matrices}
Consider an arbitrary  matrix $X$ of order $N$. Using the standard singular
value decomposition we write $X=UEV^{\dagger}$, where the diagonal matrix $E$
contains singular values $X$, equal to square roots of eigenvalues of the
positive matrix $\sigma_i(X)=\sqrt{\lambda_i(XX^{\dagger})}$, with
$i=1,\dots,N$. Unitary matrices $U$ and $V$ are formed by eigenvectors of
Hermitian matrices, $XX^{\dagger}$ and $X^{\dagger}X$, respectively. The
spectral norm of $X$ is defined by its largest singular value,
$||X||_{sp}=\sigma_{\rm max}(X)$.

One can rescale $X$ by its operator norm and arrive at the matrix
$X_0=X/\sigma_{\rm max}$ with  all singular values belonging to the unit
interval  $I$. To get a direct analogue of the matrix map (\ref{log2}) one can
write  $X_1=a U[E({\mathbb I}-E)]V^\dagger$, where $U,V$ and $E$ are determined
by the singular value decomposition, $X_0=UEV^\dagger$. The bases fixed by the
unitaries $U$ and $V$ remain invariant under the time evolution, hence the
proposed variant of the logistic map for an arbitrary normalized, non-Hermitian
matrix $X_0=UEV^{\dagger}$ reads,
\begin{equation}
X_{t+1}=f(X_t) \ := \ a \bigl[  X_{t} ({\mathbb I} - VU^\dagger X_t) \bigr].
\label{log4}
\end{equation}
It is easy to see that if the initial matrix $X$ is Hermitian, it can be
diagonalized unitarily,  $X=UEU^{\dagger}$, so $U^\dagger=V^{\dagger}$ and
Eq.(\ref{log4}) reduces to Eq.(\ref{log2}).

In the general, non-Hermitian case, both unitary matrices $U$ and $V$ of
eigenvectors are fixed during the time evolution, as the map (\ref{log4}) acts
only on the singular values of $X$. Their dynamics is then governed by the
scalar logistic map (\ref{log1}), so we obtain a statement analogous to
Proposition~\ref{prop:1}.

\begin{proposition}\label{prop:2}
Consider an arbitrary ensemble of random (non--Hermitian) matrices of order $N$
with the density of singular values described by any  continuous function
supported in the unit interval. Iterating its elements  $t$ times by the matrix map
(\ref{log4}) one obtains asymptotically an ensemble with the density of
singular values converging in the limit $t \to \infty$ to the invariant measure
$\mu_a^*$ of the logistic map (\ref{log1}).
\end{proposition}

Evolution of the asymptotic density $P(x)$ of singular values of rescaled random
Ginibre matrices \cite{Me04}  of size $N=1000$ under the action of logistic map
(\ref{log4}) with $a=3.9$ is presented in Fig.~\ref{fig:iterate-ginibre}.
Observe a fast convergence of the initial quarter-circle law to the fractal
invariant measure $\mu_{3.9}^*$.

The method proposed above allows us to generate novel ensembles of
random matrices with fractal distribution of singular values related to the 
invariant measure of the logistic map. 
Interestingly, in some cases of the logistic parameter $a$ 
one can find analytic expressions
 for the asymptotic density of eigenvalues in the complex plane.

 Derivation of  the  eigenvalue distribution of matrices of the form,
$X=UP$, where $P$ is a positive matrix with known
distribution of eigenvalues and $U$ is a Haar random unitary,
based on the theorem of  Haagerup and Larsen~\cite{HL00}
is presented in  Appendix  \ref{non_normal}.
This technique was used to analyze
 spectral properties of the ensembles
of random matrices  generated by the matrix logistic map
(\ref{log4}), which depend on the value of the
logistic parameter $a$ -- see Appendix \ref{spectra}.

In the case of the fixed point logistic dynamics, $a<3$,
the single Dirac delta of the distribution
of singular values leads to
an ensemble of matrices unitary up to rescaling,
with  eigenvalues forming a circle in the complex plane.
Increasing the logistic parameter,
$3< a < a_c \approx 3.569$,
 one enters the dynamics with period two.
 In this case  the distribution of singular values,
 formed by a superposition of two Dirac deltas,  
leads to a {\sl single ring}
of complex eigenvalues ~\cite{FZ97,GKZ11},
with the minimal and maximal radius
known analytically - see Appendix \ref{app2}.
Also the case of fully chaotic dynamics, $a=4$,
can be treated analytically, and the
corresponding spectral distribution is derived in 
Appendix \ref{app3}.
  
\section{Coupled logistic map related to a graph}
We shall consider here a more general form of Eq.~\eqref{log2}, in which the
scalar parameter $a$ is replaced by a fixed positive matrix $A=BB^{\dagger}$, so
the manifestly positive, {\sl matrix logistic map} reads,
\begin{equation}
\label{logi_graph}
H_{t+1}=f_B(H_t)=B H_t(\1 - H_t) B^\dagger.
\end{equation}
In this setup we easily recover Eq.~\eqref{log2} by setting $B=\sqrt{a} \1$.
The simplest case of the model with matrices $B$ of size $N=2$
is discussed in  Appendix \ref{inv-coupled}.

\begin{figure}[!h]
\centering\includegraphics[width=\columnwidth]{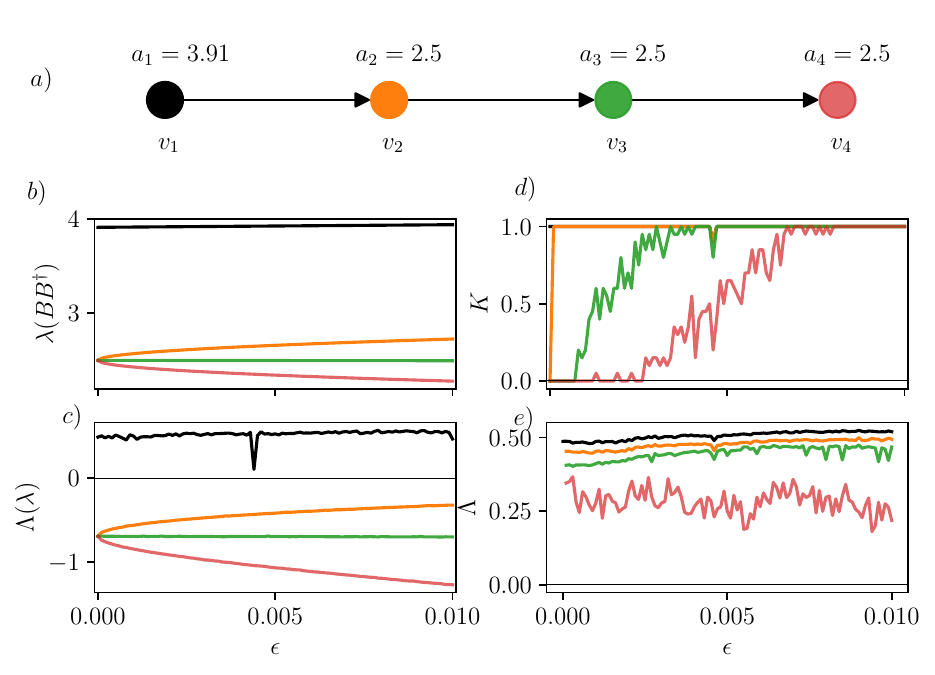} 
  
\caption{Coupled logistic map corresponding  to a linear, four-vertex directed
graph -- see panel a). Uncoupled system exhibits chaotic dynamics at the black
vertex $v_1$ determined by the parameter $a_1=3.91$ and regular dynamics
($a_i=2.5$) at color vertices $v_2,v_3,v_4$. Quantities describing the system
plotted as a function of the coupling strength $\varepsilon$: b) ordered
eigenvalues $\lambda_i$ of the coupling matrix $A=BB^{\dagger}$; c) Lyapunov
exponents $\Lambda_i(a)$ corresponding to the uncoupled logistic map
(\ref{log1}) with parameter $a=\lambda_i$; d) value $K$ of the $0-1$ chaos test;
e) Four Lyapunov exponents $\Lambda_i$ for the entire $N=4$ dimensional
dynamical system (\ref{logi_graph}). The further the vertex $v_i$ is situated
from the chaotic dynamics at vertex $v_1$, the larger value of the coupling
parameter $\varepsilon$ is necessary to generate chaotic dynamics at $v_i$.}
\label{fig:linear}
\end{figure}

\begin{figure}[!h]
\centering\includegraphics[width=\columnwidth]{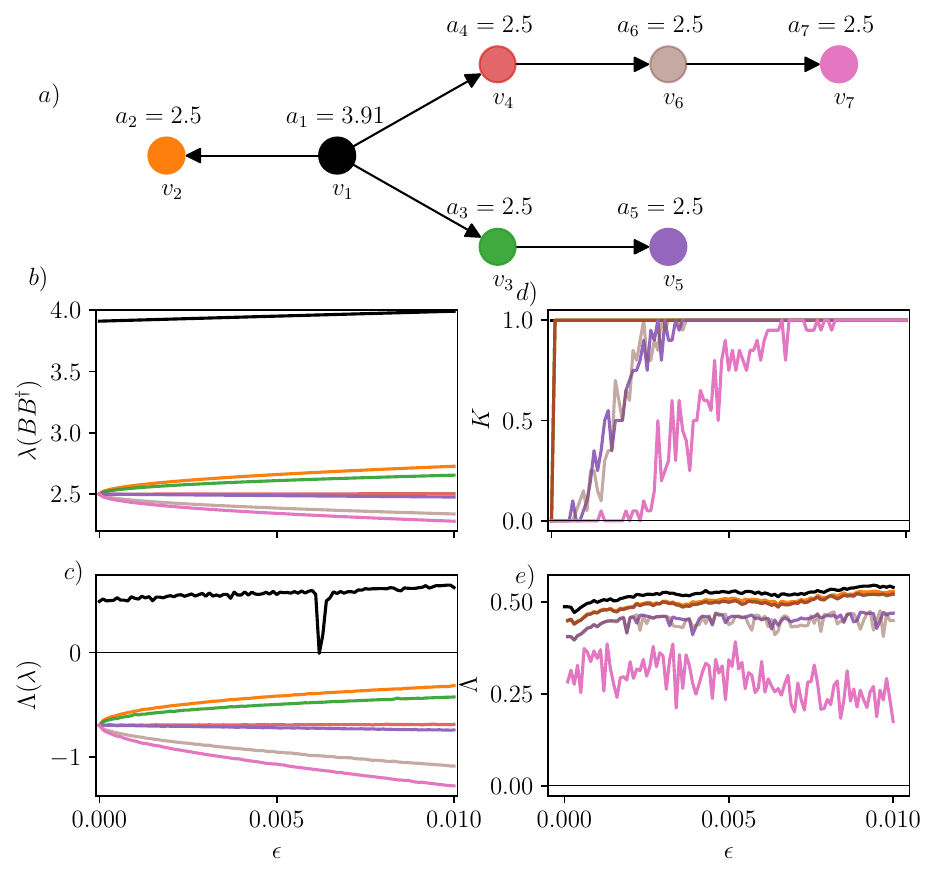} 

\caption{Coupled logistic maps on a seven-vertex star graph a) with chaotic
dynamics, at the central vertex $v_1$ (in black) determined by logistic
parameters $a_1=3.91$ and regular dynamics at other vertices $v_i$ with
$a_i=2.5$, for $ i=2,\dots 7$. Changes of the dynamics with coupling parameter
$\varepsilon$ for $N=7$ dimensional system shown in panels b)-e) presented as in
Fig.~\ref{fig:linear} (see the caption for a detailed explanation of subfigures
presented here). The transfer to chaos for vertices $v_2,v_3$ and $v_4$ directly
coupled with $v_1$ occurs for the same value of $\varepsilon$. A stronger
coupling is needed to induce chaotic dynamics at vertices $v_5$ and $v_6$.
Dynamics at vertex $v_7$, more distant from the source of chaos $v_1$, becomes
chaotic for even larger values of $\varepsilon$.}
\label{fig:star}
\end{figure}

To demonstrate key features of this model we study matrix $B$ with the structure
determined by the adjacency matrix of any directed graph with $N$ vertices. The
first one is a linear graph, presented in Fig.~\ref{fig:linear}a and the other
one has a structure of the star graph with multiple layers shown in
Fig.~\ref{fig:star}a. In the first case of the graph with four vertices the
matrix %$B=B_4$ 
reads,
\begin{equation}
  B=B_4 = \begin{pmatrix}
    \sqrt{a_0} & \sqrt{\varepsilon} & 0 & 0 \\
    0 & \sqrt{a_1} & \sqrt{\varepsilon} & 0 \\
    0 & 0 & \sqrt{a_2} & \sqrt{\varepsilon} \\
    0 & 0 & 0 & \sqrt{a_3}
  \end{pmatrix} ,
  \label{B4}
\end{equation}
where $\varepsilon \ge 0$ is a coupling parameter. It is crucial to ensure that
the eigenvalues $\lambda$ of $A=B_4B_4^\dagger$ belong to the interval $[0, 4]$.
This constraint prevents the iterates from diverging outside the range $[0, 1]$,
even in the presence of coupling. For example, let us consider the extreme case
where the parameters $a_0$, $a_1$, $a_2$, and $a_3$ are all equal to four 4 and
$\varepsilon >0$. Then, starting with $D = \frac12 \Id$ can lead to values
outside the desired interval. By limiting the eigenvalues, we ensure that the
system remains bounded and well-behaved throughout the iterations.

Models described by these coupling matrices $B$ can be viewed as coupled
logistic maps with interaction determined by a graph. We analyzed both systems
numerically, in each case performing  two tests for chaos. First, we calculate
the Lyapunov characteristic exponents (LCE) $\Lambda_i$ associated with dynamics
$x_i$ at the vertices $v_i$ of  the graphs, making use of the numerical
techniques described in \cite{BGGS80, Ka94, Sko10}. Secondly, we perform the $0-1$
test for chaos~\cite{gottwald20160}. In both cases we set $a_1=3.91$ and
$a_i=2.5$ for all $i>1$, which imply chaotic dynamics at  $v_1$ and regular one
at all other vertices $v_i$ -- see Fig.\ref{fig:bifur}.

Results obtained for a positive value $\varepsilon$ confirm the transfer of
chaotic dynamics between the vertices of the graph due to the coupling between
subsystems. Increasing the constant $\varepsilon$, which couples the chaotic
dynamics at the vertex $v_1$ with other vertices, one observes that the
transition to chaos occurs first at the nearest neighbor $v_2$, while subsequent
vertices $v_3$ and $v_4$ display chaos only for significantly larger values of
$\varepsilon$. 

To further drive our point about the transfer of chaos, we present numerical
results for the star-shaped graph shown in Fig.~\ref{fig:star}. As in the case
of a linear graph we see the chaotic behavior appearing as a function of the
distance from the chaotic vertex. Note that for two vertices with the same
distance from the chaotic vertex $v_1$, the transfer to chaos occurs for the
same value of the coupling $\varepsilon$.

\section{Concluding Remarks}
 In this work we introduced a matrix version of the celebrated logistic map,
$x'=ax(1-x)$, fundamental in studies of chaotic dynamics \cite{St00,AD06}. In
the first step we allowed the variable $x$ to be replaced by a Hermitian matrix
$H$ of a fixed size $N$ satisfying $0 \le H \le {\mathbb I}$. This assumption
leads to a significantly generalized version (\ref{log2}) of the model  and
introduces novel ensembles of random matrices, which asymptotically (in size $N$
and time $t$) display singular fractal level distributions. In a more general
set-up of matrix logistic map acting on non-Hermitian matrices (\ref{log4}) the
eigenvalues of the asymptotic ensemble form a ring in the complex plane
consistent with the `single ring theorem'~\cite{FZ97,GKZ11}, even though the
asymptotic distribution of the singular values is fractal -- see
Fig.~\ref{fig:herm38}. A more detailed discussion of the emergence of the
single ring is presented in Appendix~\ref{non_normal}.

In the second step we assumed that the logistic parameter $a$ is replaced by a
positive matrix $A=BB^{\dagger}$ of a fixed order $N$. Defining $B$ to be the
adjacency matrix of a given directed graph $\Gamma$ with $N$ vertices we find a
related logistic map (\ref{logi_graph}) acting on matrices of order $N$. Each
vertex of $\Gamma$ represents a scalar logistic map, while the interactions
between the maps are determined by the edges of $\Gamma$. In this way we arrive
at a versatile tool to analyze dynamics in complex networks and demonstrate
transfer of chaos between subsystems corresponding to particular vertices of the
graph. The proposed matrix model is wide enough that it can be also used to mimic
nonlinear quantum processes, as the iterated positive matrix can be interpreted
as a quantum state subjected to nonlinear evolution.

\section*{Acknowledgements}
It is a pleasure to thank Piotr Gawron, Wojciech S{\l}omczy{\'n}ski and Tomasz
Szarek for discussions on invariant measures of dynamical systems,
as well as %to 
Maciej
Nowak, Zbigniew Pucha{\l}a and Kamil Szpojankowski for valuable consultations on
spectra of non-Hermitian random matrices
 and  Adam Bollt, 
 Oskar Pro{\'s}niak and Piotr Staro{\'n} for stimulating interactions. Financial
support by Narodowe Centrum Nauki under the Quantera project number
2021/03/Y/ST2/00193 and by Foundation for Polish Science under the Team-Net
project no. POIR.04.04.00-00-17C1/18-00
 is gratefully acknowledged.

% \clearpage
\appendix

\section{Asymptotic eigenvalues of random non-normal matrices}
\label{non_normal}

In this Appendix we consider the limiting eigenvalue distribution of matrices
$X=UP$, where $P$ is a positive matrix with limiting eigenvalue distribution
$\nu_P(x)$ and $U$ is a Haar random unitary.

In order to study such matrices, we recall some facts about the S-transform,
which was introduced in free random probability by Voiculescu~\cite{Vo89}. Let
us consider a Hermitian matrix $H$. We define its Cauchy transform (Green's
function) as
\begin{equation}
G_H(z) = \lim_{N \to \infty}\frac1N \langle \Tr(z\1-H)^{-1} \rangle = \int 
\frac{\nu_H(x)}{z-x} \dd x.
\end{equation}

The Green's function is related to moments $m_{Hk}$ of $\nu_H$
\begin{equation}
G_H(z) = \sum_{k=0}^\infty \frac{m_{Hk}}{z^{k+1}}.
\end{equation}
A more convenient generating function, given by a power series in $z$ rather 
than in $1/z$ is
\begin{equation}
\psi_H(z) = \frac1z G_H\left(\frac1z\right) - 1 = \sum_{k=1}^\infty m_{Hk} z^k.
\end{equation}
Let us denote by $\chi_H(z)$ the inverse of $\psi_H(z)$
\begin{equation}
\chi_H(\psi_H(z))=\psi_H(\chi_H(z))=z,
\end{equation}
which can be expressed as a power series in $z$ provided that $m_{H1} \neq 0$. 
The S-transform  \cite{BJN11}
of $H$ is related to $\chi_H$ as
\begin{equation}
S_H(z)=\frac{1+z}{z}\chi_H(z).
\end{equation}

Now we can recall the seminal result by Haagerup and Larsen~\cite{HL00}, which
implies that the limiting eigenvalue distribution of $X$, $\varrho_X(z,
\overline{z})$, is radially symmetric and related to the S-transform of $P^2$,
\begin{equation}
S_{P^2}(F_X(r)-1)=\frac{1}{r^2},
\end{equation}
where $F_X(r)$ is the radial cumulative density function of the eigenvalues of 
$X$
\begin{equation}
F_X(r) = \int_{|z|\leq r} \varrho_X(z, \overline{z}) \dd^2z = 2\pi \int_0^r
s\nu_X(s)\dd s= \int_0^r p_X(s) \dd s.
\label{eq:int-f}
\end{equation}
It is related to the eigenvalue density $\varrho_X(z,\overline{z})=\mu_X(|z|)$.
The integrand $p_X(s)\dd s = 2 \pi s \nu_X(|z|)$ can be interpreted as the
probability of finding the eigenvalues of $X$ in a ring of radii $|z|$ and
$|z|+\dd |z|$,
\begin{equation}
p_X(r) = \partial_r F_X(r).\label{eq:px-on-f}
\end{equation}
Furthermore, the Haagerup-Larsen theorem tells us that the support of the 
eigenvalue density of $X$ is a single ring bounded by circles of radii
\begin{equation}
R^2_{\rm max} = \int_0^\infty x\nu_{P^2}(x) \dd x, \; \; R^{-2}_{\rm min} = 
\int_0^\infty x^{-1} \nu_{P^2}(x) \dd x.
\end{equation}
%Now we have all the ingredients to study the stationary eigenvalue densities of 
%non-normal matrices evolved according to the matrix 
%logistic map~(5).

\section{Spectral densities of ensembles of nonHermitian random matrices
determined by logistic map}
\label{spectra}

We are going to apply the techniques presented in
Appendix  \ref{non_normal}
to study asymptotic distribution of eigenvalues
of non-Hermitian random matrices
generated by the matrix 
logistic map~(\ref{log4}).
The structure of the spectra depends on the value of the logistic parameter $a$.

\subsection{Fixed point dynamics for logistic parameter $a<3$}
\label{app1}
In this regime there is a single stationary value for the logistic map, $x_c$. 
Hence $\nu_P(x) = \delta(x-x_c)$ and $X=x_c U$ and the eigenvalue density is 
supported on a circle of radius $x_c$
\begin{equation}
p_X(r) = \delta(r-x_c).
\end{equation}

\begin{figure}[!h]
  \centering\includegraphics[width=0.95\columnwidth]{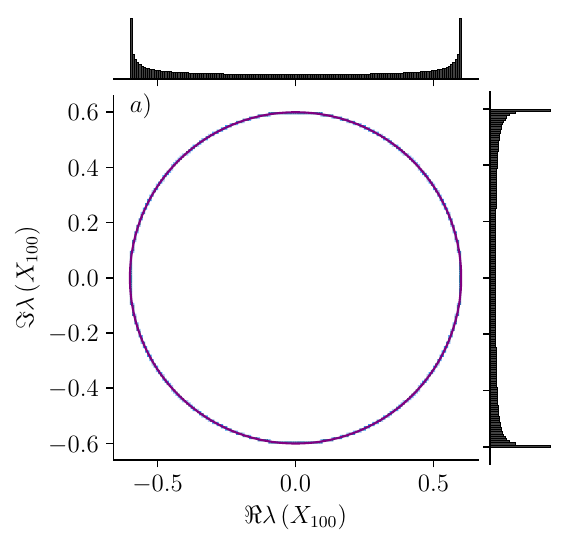} 
  
  \caption{Distribution of eigenvalues of matrix $X_{100}$ obtained by iterating
  $100$ times a random  Ginibre matrix $X_0$ of dimension $N=1000$ by the matrix
  logistic in the case of a single fixed-point, $a=2.5$. The side plots are the
  marginal distributions of the real and imaginary parts. The distribution is
  supported on a circle of radius .}\label{fig:circle}
\end{figure}

An example of this distribution along with marginal distributions are shown in
Fig.~\ref{fig:circle}. This was obtained by iterating 1000 initial matrices of
size $N=1000$ for 100 steps and obtaining the final eigenvalues.

\subsection{Intermediate dynamics for logistic parameter $3\leq a \leq a_c$}
\label{app2}
In the parameter range, for
$a < a_c \approx 3.569$, %95$ 
 the dynamics is periodic and a sequence of period doubling effects occur while
 increasing $a$. In this case the logistic map~(1) has exactly two fixed points,
 the invariant measure is formed by a combination of two Dirac delta functions,
 $\nu_P(x) = \frac12 \delta(x-\alpha_+)+\frac12 \delta(x-\alpha_-)$ and
 $\alpha_+>\alpha_->0$. Assume than that the singular values of a non-Hermitian
 matrix $X=UP$ is given by such a combination of Dirac deltas. The resulting
 radial density $p_X(r)$ of complex eigenvalues of $X$ reads
\begin{equation}
p_X(r) = r\left( \frac{\alpha_+^2}{(\alpha_+^2-r^2)^2} + \frac{\alpha_-^2}{(\alpha_-^2-r^2)^2} \right).
\label{eq:dist-10-3}
\end{equation}
This distribution is obtained by following the steps described in the previous section
and calculating an explicit form of Eq.~\eqref{eq:px-on-f}.
  
The support is a disc of radii
\begin{equation}
r_{\rm min} = \frac{\sqrt{2}\alpha_+ \alpha_-}{\sqrt{\alpha_+^2+\alpha_-^2}}, \; \; r_{\rm max} = 
\frac{\sqrt{\alpha_+^2+\alpha_-^2}}{\sqrt{2}}.
\end{equation}

Let us look at a specific example. Set $=10/3$, then 
$\alpha_+=\frac{13+\sqrt{13}}{20}$, $\alpha_-=\frac{13-\sqrt{13}}{20}$. We get
\begin{equation}
r_{\rm min} = \frac35 \sqrt{\frac{13}{14}} \approx 0.578, \; \; r_{\rm max} = 
\frac{1}{10}\sqrt{\frac{91}{2}} \approx 0.675.
\label{rmin_rmax}
\end{equation}

\begin{figure}[!h]
\centering\includegraphics[width=0.95\columnwidth]{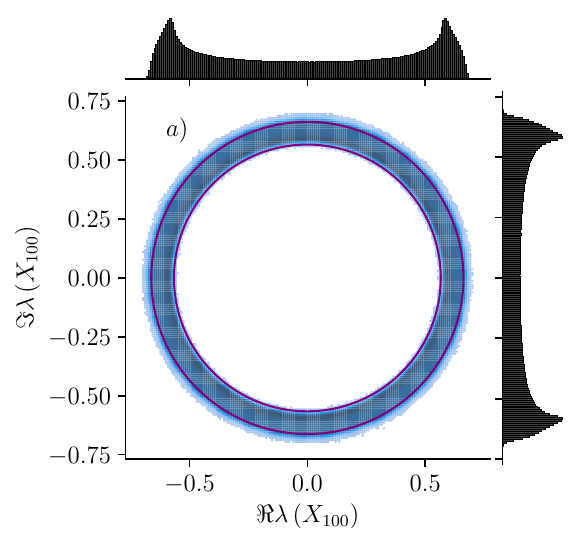}\\
\includegraphics[width=\columnwidth]{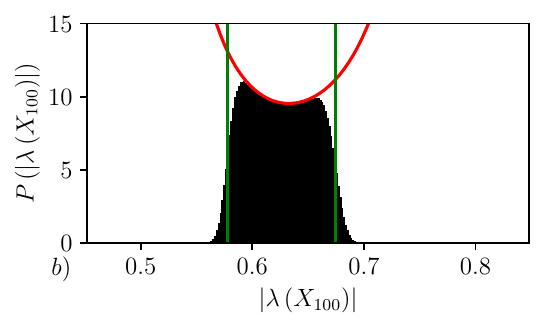}
  
\caption{(top) Distribution of eigenvalues of matrix $X_{100}$ obtained by
iterating $100$ times a random  Ginibre matrix $X_0$ of dimension $N=1000$ by
the matrix logistic map with $a=10/3$, two fixed-points. The side plots are the
marginal distributions of the real and imaginary parts. The circles mark the
edges of the disc of the radii from~\eqref{rmin_rmax}. (bottom) Plot showing the
distribution of the absolute values of the eigenvalues of $X_{100}$. The green
lines mark the support of the asymptotic distribution, the red line represents
distribution~\eqref{eq:dist-10-3}.}\label{fig:disc}
\end{figure}

Again, we show of this distribution in Fig.~\ref{fig:disc}. As before, this
was obtained by iterating 1000 initial matrices of size $N=1000$ for 100 steps
and obtaining the final eigenvalues.

\subsection{Chaotic dynamics for logistic parameter $a=4$}
\label{app3}
\begin{figure}[!ht]
  \centering\includegraphics[width=0.95\columnwidth]{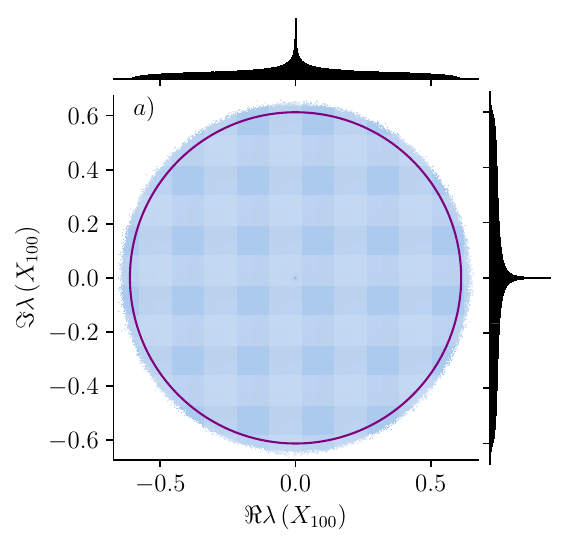}\\
  \includegraphics[width=\columnwidth]{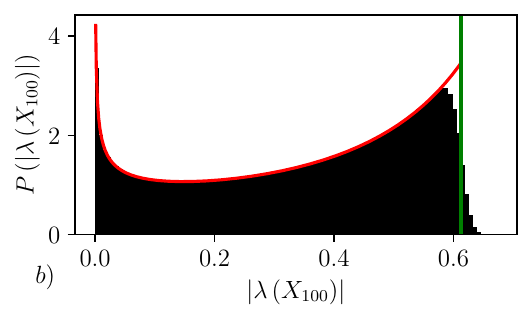}
    
  \caption{(top) Distribution of eigenvalues of matrix $X_{100}$ obtained by
  iterating $100$ times a random Ginibre matrix $X_0$ of dimension $N=1000$ by the
  matrix logistic map in the fully chaotic case, $a=4$. The side plots are the
  marginal distributions of the real and imaginary parts. The circles mark the
  edges of the disc of the radii from~\eqref{rmin_rmax}. (bottom) Plot showing the
  distribution of the absolute values of the eigenvalues of $X_{100}$. The green
  line marks the support of the asymptotic distribution, the red line represents
  distribution~\eqref{eq:arcsin2}.}\label{fig:full-disc}
\end{figure}

In the fully chaotic case, $a=4$ the invariant measure  $\nu_P(x)$ is  given by
the arcsine law~(2) -- see e.g. \cite{Ott}. Assuming that the
singular values of a non-Hermitian matrix $X$ are distributed according to this
law, we apply now  Eq. \eqref{eq:int-f} and \eqref{eq:px-on-f} to derive the resulting radial density
$p_X(r)$. It is supported on $[0, \sqrt{3/ 8}]$ and given by a derivative,

\begin{equation}
\begin{split}
p_X(r) =& \frac{\partial}{\partial r} \left[ -\frac{r^2+1}{4 \left(r^2-1\right)}
 +\frac{1}{2} \sqrt{T(r)+\frac{Q(r)}{W(r)}}+\right.\\
 &\left. -\frac{1}{2}
\sqrt{2T(r)+\frac{r^2+1}{4 \left(1-r^2\right)
\sqrt{T(r)+\frac{Q(r)}{W(r)}}}-\frac{Q(r)}{W(r)}} \right],
\label{eq:arcsin2}
\end{split}
\end{equation}
with
\begin{equation}
\begin{split}
T(r) =& \frac{3 r^4-2 r^2+3}{12 \left(r^2-1\right)^2},\\
Q(r) = & 4r^2(3 r^4 - 5 r^2 + 3)+\left[U(r)\right]^{2/3},\\
W(r)=&6(r^2 - 1)^2 \left[ U(r) \right]^{1/3},\\
U(r) =& R(r) + 27 r^{10}-72 r^8+98 r^6-72 r^4+27 r^2,\\
R(r) =& 3 \sqrt{3} \sqrt{r^4 \left(r^2-1\right)^6 \left(27 r^4-46
r^2+27\right)}.
\label{eq:arcsin3}
\end{split}
\end{equation}
This asymptotic probability distribution describes well numerical results for
the radial distribution of complex eigenvalues obtained by iteration of random
Ginibre (non-Hermitian) matrices of size $N=1000$ -- see
Fig.~\ref{fig:full-disc}.

\section{$N=2$: Invariant measures for two matrix-coupled logistic maps}
\label{inv-coupled}

\begin{figure}[!htp]
\centering\includegraphics{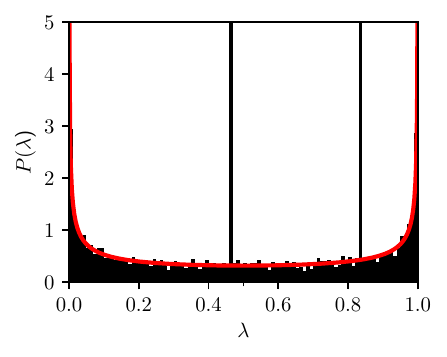} 

\caption{Invariant measure of two decoupled logistic maps, obtained by setting
$\varepsilon=0$ in Eq.~\eqref{logi_graph} for $N=2$ with parameters $a_1=10/3$
and $a_2=4$ forms a combination of both invariant measures of the logistic map
(1): superposition of two Dirac measures $\delta(x-x_\pm)$ localized at  
$x_\pm=\frac{13 \pm \sqrt{13}}{20}$ corresponding to  $a=a_1$ and the continuous
arcsine distribution (2) obtained for $a=a_2$.}
\label{fig:convex}
\end{figure}

In this Appendix we analyze the matrix logistic map in the simplest case of
$N=2$. Consider first the map (\ref{logi_graph}) with a diagonal matrix
$B=\diag(\sqrt{a_1}, \sqrt{a_2})$. This assumption corresponds to two decoupled
logistic maps. The resulting steady state distribution is a convex combination
of the steady state distributions for parameters $a_1$ and $a_2$. Such an
example is provided  in Fig.~\ref{fig:convex}.

\begin{figure}[!h]
\centering\includegraphics{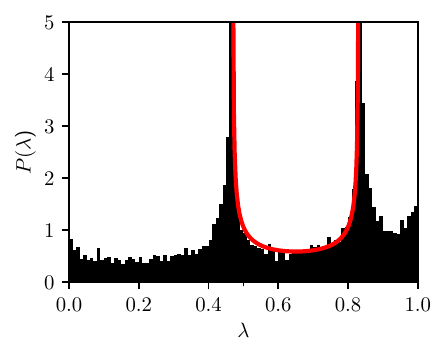}

\caption{Exemplary coupled matrix logistic maps with $N=2$ subsystems. The
coupling is achieved by setting  to Eq.~\eqref{logi_graph} matrix $B$ with
parameters, $a_1=a_2 = 1/ \sqrt{60 + 11 \sqrt{30}}$ and $\varepsilon=(5/3)^{1/4}
2^{3/4}$. This choice implies that  the distinct eigenvalues of $BB^\dagger$
read $\lambda_1=10/3$ and $\lambda_2=4$. Observed singularities of the invariant
measure at $x_\pm= (13 \pm \sqrt{13})/{20}$ correspond to fixed points of the
logistic map~(1) with  $a=\lambda_1=10/3$. Red curve represents the
rescaled arcsine distribution~(2) which approximates the part of the
distribution between both singularities.}
\label{fig:coupled}
\end{figure}

The second example features a richer behavior two logistic maps coupled by a
non-diagonal matrix $B$. The coupling is achieved using the matrix
$B=\begin{pmatrix} \sqrt{2 \sqrt{10/3}} & \sqrt{2/3 (11 - 2 \sqrt{30})}\\ 0 &
\sqrt{2 \sqrt{10/3}} \end{pmatrix}$. This gives us the eigenvalues of
$A=BB^\dagger$ equal to $\lambda_1=10/3$ and  $\lambda_2=4$. The resulting
steady-state measure is shown in Fig.~\ref{fig:coupled}. Note that due to the
positive coupling parameter $\varepsilon >0$ the invariant density differs from
the combination of invariant densities for the maps with parameters
determined by both eigenvalues of $A$, i.e.  $a_1=\lambda_1=10/3$ and
$a_2=\lambda_2=4$.

\end{document}